\documentclass{elsart}

\usepackage{natbib}

\usepackage{graphicx}
\usepackage{epstopdf}
\usepackage{enumerate}
\usepackage{amsmath}
\usepackage{amssymb}
\usepackage{pslatex}

\usepackage{amssymb}

\usepackage[latin1]{inputenc}
\begin{document}

\begin{frontmatter}



\title{Emergence and resilience of cooperation in the spatial 
Prisoner's Dilemma via a reward mechanism}


\author[Est,USB]{Ra\'ul Jim\'enez},
 \ead{rjjimene@est-econ.uc3m.es}
\author[Eco,USB]{Haydee Lugo} \ead{haydee.lugo@cantv.net} 
\author[Mat]{Jos\'e A.\ Cuesta} \ead{cuesta@math.uc3m.es}
\author[Mat,bifi]{Angel S\'anchez\corauthref{cor}} \corauth[cor]{Corresponding author.}\ead{anxo@math.uc3m.es}
\address[Est]{Departamento de Estad\'\i stica, Facultad de Ciencias Sociales y 
Jur\'\i dicas, Universidad Carlos III de Madrid, 28903 Getafe, Spain}
\address[Eco]{Departamento de Econom\'\i a, Facultad de Ciencias Sociales y 
Jur\'\i dicas, Universidad Carlos III de Madrid, 28903 Getafe, Spain}
\address[Mat]{Grupo Interdisciplinar de Sistemas Complejos (GISC), 
Departamento de Matem\'aticas, Escuela Polit\'ecnica Superior, 
Universidad Carlos III de Madrid, 28911 Legan\'es, Spain}
\address[USB]{Departamento de C\'omputo Cient\'\i fico y Estad\'\i stica, 
Universidad Sim\'on Bolivar, A.\ P.\ 89000, Caracas 1090, Venezuela}
\address[bifi]{Instituto de Biocomputación y Física de Sistemas Complejos,
Universidad de Zaragoza, 50009 Zaragoza, Spain}

\begin{abstract}
We study the problem of the emergence of cooperation in the spatial 
Prisoner's Dilemma. The pioneering work by \citet{nowak-may}
showed that large initial populations 
of cooperators can survive and sustain cooperation
in a square lattice with imitate-the-best evolutionary dynamics. 
We revisit this problem 
in a cost-benefit formulation suitable for a number of biological applications.
We show that if a 
fixed-amount reward is established for cooperators to share, a single
cooperator can invade
a population of defectors and form structures that are resilient to 
re-invasion even if the reward mechanism is turned off. We discuss 
analytically the case of the invasion by a single cooperator and present
agent-based simulations for small initial fractions of cooperators.
Large cooperation levels, in the sustainability range, are found. 
In the conclusions we discuss possible applications of this model as well
as its connections with other 
mechanisms proposed to promote the emergence of cooperation. 
\end{abstract}


\end{frontmatter}

\section{Introduction}
\label{intro}

The emergence of cooperative behavior among unrelated individuals is one of the
most prominent unsolved problems of current research \citep{pennisi}. While such
non-kin cooperation is evident in human societies \citep{hammerstein}, it is by
no means exclusive of them, and can be observed in many different species \citep{doebeli} 
down to the level of microorganisms \citep{velicer,levin}. This conundrum can be 
suitably formulated in terms of evolutionary game theory \citep{maynard,gintis,nowak-sigmund,nowak-book} by studying games that are 
stylized versions of social dilemmas \citep{dilemas}, e.g., situations in
which individually reasonable behavior leads to a situation in which 
everyone is worse off than they might have been otherwise. Paradigmatic
examples of these dilemmas are the provision of public goods \citep{samuelson}, 
the tragedy
of the commons \citep{hardin}, and the Prisoner's Dilemma (PD) \citep{axelrod}. The 
first two of them involve multiple actors, while the latter involves only 
two actors, this last case being the setting of choice for a majority of models 
on the evolution of cooperation. 

The PD embodies a stringent form of social dilemma, namely a
situation in which individuals can benefit from mutual cooperation but they
can do even better by exploiting cooperation of others. To be specific, the
two players in the PD can adopt either one of two strategies:
cooperate (C) or defect (D). Cooperation results in a benefit $b$ to 
the opposing player, but incurs a cost $c$ to the cooperator (where
$b>c>0$). Defection has no costs and produces no benefits. Therefore,
if the opponent plays C, a player gets the payoff $b-c$ if she also 
plays C, but she can do even better and get $b$ if she plays D. On 
the other hand, if the opponent plays D, a player gets the lowest payoff
$-c$ if she plays C, and it gets 0 if she also defects. In either case, 
it is better for both players 
to play D, in spite of the fact that mutual cooperation would yield higher
benefits for them, hence the dilemma. 

Conflicting situations that can be described by the PD, 
either at the level of individuals or at the level of populations
are ubiquitous. Thus, \citet{turner} showed that interactions
between RNA phages co-infecting bacteria are governed by a PD. {\em 
Escherichia coli} stationary phase GASP mutants in starved cultures 
are another example of 
this dilemma \citep{vulic}.  A PD
also arises when different yeasts compete by switching from respiration 
to respirofermentation when resources are limited \citep{frick}. 
Hermaphroditic fish that alternately release sperm and eggs end up 
involved in a PD with cheaters that release only sperm with less 
metabolic effort \citep{dugatkin}.  A recent study of cooperative territorial defence in lions ({\em Panthera leo}), described the correct ranking structure for a PD \citep{legge}. And, of course, the PD applies to very many different 
situations of interactions between human individuals or collectives
\citep{axelbook,camerer}.

In view of its wide applicability, the PD is a suitable context to pose 
the question of the emergence of cooperation. How do cooperative individuals
or populations survive or even thrive in the context of a PD, where 
defecting is the only evolutionarily stable strategy \citep{maynard,nowak-book}?
Several answers to this puzzle have been put forward \citep{nowak-review}
among which the most relevant examples are kin selection theory 
\citep{hamilton}, reciprocal altruism or direct reciprocity \citep{trivers,%
axelrod}, 
indirect reciprocity \citep{nowak-sigmund2}, emergence of cooperation
through punishment \citep{fehr} or the existence of a spatial or social 
structure of interactions \citep{nowak-may}. This last approach has 
received a great deal of attention in the last decade and has proven 
a source of important insights into the evolution of cooperation (see
\citet{szabo} for a recent and comprehensive review). One such insight
is the fact that cooperators can outcompete defectors by forming clusters
where they help each other. This result, in turn, leaves open the question 
of the emergence of cooperation in a population with a majority of defectors. 
Recently, it has been shown \citep{japo} that, if the average number 
of connections in
the interaction network is $k$, the condition $b/c>k$ implies that 
selection favors
cooperators invading defectors in the weak selection limit,  i.e., when 
the contribution of the game to the fitness of the individual is very 
small. However, a general result valid for any intensity of the selection
is still lacking. 

In this paper, we propose a new mechanism for the emergence of cooperation,
which we call {\em shared reward}. In this setting, players interact through
a standard PD, but in a second stage cooperators receive an additional 
payoff coming from a resource available only to them and not to defectors. 
It should be emphasized that similar reward mechanisms may be relevant for
a number of specific applications, such as mutualistic situations with
selection imposed by hosts rewarding cooperation or punishing less 
cooperative behavior (see, e.g., \citet{denison} and references therein). 
Another context that may be modelled by our approach is team formation 
in animal societies \citep{anderson}, e.g., in cooperative hunting
\citep{packer}. On the other hand, the idea of a shared reward 
could be implemented
in practice as a way to promote cooperation in human groups or, 
alternatively, may arise from costly signaling prior to the game,
when the exchange of cooperative signals among cooperators is free
\citep{skyrms}. 
As we will see, this scheme makes it possible for a single cooperator to 
invade a population of defectors. Furthermore, when strategies evolve 
by unconditional imitation \citep{nowak-may}, cooperation persists after
the additional resource has been exhausted or turned off. We present 
evidence for these conclusions coming from numerical simulations on a 
regular network. In the
conclusion, we
discuss the reason for this surprising result and the relation of our
proposal to previous work on evolutionary games on graphs and to public
goods games. 

\section{ Spatial Prisoner's dilemma with shared reward}
\label{PD}

Our model is defined by a two-stage game
{on a network}. In the first stage, players
{interact with their neighbors and gobtained payoffs as prescribed by the PD game,}
whose payoff matrix in a cost-benefit context is
given by 
\begin{equation}
\label{matrixPD}
\begin{array}{ccc}
 & \mbox{ }\, C & \!\!\!\!\!\! D \\
\begin{array}{c} C \\ D \end{array} & \left(\begin{array}{c} b-c \\b
\end{array}\right. & \left.\begin{array}{c} -c \\ 0
\end{array}\right). \end{array} \end{equation} 
Subsequently, in the second stage of the game, 
a fixed amount $\rho$ is distributed among {\em all cooperators}. 
It is important to realize at this point that such a
two-stage game is only interesting in a population setting: in a two-player game,
the second stage
would amount to shift the cooperator's payoff by {$\rho/2$ or $\rho$, depending
on the opponent's strategy}.
Then, for $\rho<c$ we would simply have another PD,
whereas for 
$2c>\rho>c$ we would have the Hawk-Dove or Snowdrift game 
\citep{maynard}, and for 
$\rho>2c$ we would have the trivial Harmony game (also called 
Byproduct Mutualism \citep{dugatkinold,connor}. In a population 
setting, the amount received by a cooperator depends on the number
of cooperators in the total population and is therefore 
subject to evolution as the population itself changes. 

In order to write down the payoffs for the game after the second
stage, we need to introduce some notation. Let us consider 
a population of $N$ players, each of whom plays the game against
$k$ other players. For player $i$, $1\leq i \leq
N$, let us denote by 
{$V_i$ the number of cooperators among the opponents of $i$},
{and by $N_c$ the total number of cooperators in the population}.
The payoffs can then be written as follows:
\begin{equation} 
\label{bb}
P_i= \left\{
\begin{array}{ll}
V_i b - kc + \displaystyle{\frac{\rho}{N_c}}, & \mbox{\rm if $i$ cooperates} \\
V_ib,& \mbox{\rm if $i$ defects}. 
\end{array} \right.
\end{equation}

This mechanism to reward cooperation has been studied by \citet{nosotros} 
in a game theoretical model of  $n$ players with no spatial structure. 
As stated above, our goal here
is to understand whether or not the mechanism of
the shared reward can explain the emergence of cooperation in the 
Prisoner's Dilemma on networks. 
To address this problem, we will consider below this 
game in the framework of a spatial 
setup following the same general lines as \citet{nowak-may} for comparison. 
We place $N$ {individuals} 
on a square lattice with periodic boundary 
conditions,
{each of whom
cooperates or defects} 
with her neighbors (4, von Neumann neighborhood). 
We have chosen this neighborhood
for the sake of simplicity in the calculation;
{results for Moore neighborhood [used, e.g., by \citet{nowak-may}] can be obtained
in a straightforward manner.}
After {receiveing their payoffs  according to (\ref{bb}), all individuals}
update their strategy synchronously for the next round, by 
imitate-the-best (also called unconditional imitation)
dynamics: they look in their neighborhood for players
whose payoff is higher than their own. If there is any, 
the player adopts the strategy that led to 
the highest payoff among them (randomly chosen in case 
of a tie). We then repeat the process and
let the simulation run until the density of cooperators in the lattice
reaches an asymptotic average value or else it becomes 0 or 1
(note that these two states, corresponding 
to full defection and full cooperation, are absorbing states of
the dynamics because there
are not mutations). 

From the work by \citet{nowak-may}, we know that if we begin the 
simulation with a sufficiently large cooperator density, then the
lattice helps sustain the cooperation level by allowing cooperators
interacting with cooperators in cluster to survive and avoid 
invasion by defectors; defectors thrive in the boundaries between
cooperator clusters. What we are interested in is in the question as
to how the large initial cooperator level required by \citet{nowak-may}
may arise; if the initial number
of cooperators is small, they cannot form clusters and full defection
is finally established. On the other hand, another relevant point is 
resilience, i.e., the resistance of the cooperator cluster to re-invasion
by defectors. In this respect, we note that while the clusters obtained
by \cite{nowak-may} did show resilience, their corresponding cooperation
level was not large. As we will see below, the mechanism we are proposing
will lead to higher cooperation levels with good resilience properties,
even for medium costs. 
To address these issues, we begin by discussing
the invasion by a single cooperator placed on the center of the lattice  (in fact,
on any site, as the periodic boundary conditions make all sites 
equivalent). This, {along with the possible scenarios
of invasion by a single defector}, will lead to a classification of the different regimes 
in terms of the cost parameter. Subsequently,
we will carry out simulations with a 
very low initial concentration of  cooperators. 

\section{Invasion by a single cooperator and resilience of cooperation}

As our strategy update rule is unconditional imitation, 
the process is fully deterministic, so we can compute
analytically the evolution of the process. Thus, for the first cooperator,
seeded at {time} $t=0$,
to transform her defector neighbors into new cooperators, it is 
immediate to see that $\rho>b+4c$; otherwise, the cooperator is 
changed into a defector and the evolution ends. If the condition is
satisfied, the four neighbors become cooperators, and we have now 
a rhomb centered on the site of the initial cooperator. In what follows,
we discuss the
generic situation in the subsequent evolution of the system. 

After the initial cooperator has given rise to a rhomb, there will 
always be four types of players in the system: 
\begin{itemize}
\item The cooperators in the bulk, that interact with another four cooperators. 
\item The cooperators in the boundary, defined
as the set of cooperators that have links with defectors. These
boundary players have two cooperator neighbors or only one if they
are at the corners of the rhomb, but the key point is that they 
are always connected to a cooperator that interacts only with 
cooperators. 
\item The defectors in the boundary, that interact with one
(opposite to the corners of the rhomb) or two cooperators. 
\item The defectors in the bulk, that interact with another four 
defectors. 
\end{itemize}
For the rhomb to grow two conditions must be met:
first of all, the payoff obtained by 
{the boundary cooperators at the corner ($b-4c$ plus the reward contribution)
has to be larger than that of the boundary defectors with only one cooperator ($b$);
secondly, the payoff obtained by cooperators that have two cooperator neighbors
($2b-4c$ plus the reward contribution)
has to be larger than that of the boundary defectors that interact with two cooperators
($2b$).}
If both conditions are verified, defectors are forced 
to become cooperators by imitation.
Therefore,
we must have
\begin{equation}
\label{condicion2}
b-4c + \frac{\rho}{N_c(t)}>b \ \ \mbox{and}\ \ 2b- 4c + \frac{\rho}{N_c(t)}>2b
\iff \frac{\rho}{N_c(t)} > 4c.
\end{equation}
We thus find that the condition for invasion does not depend on the
benefit $b$. In addition, it predicts that invasion proceeds until
the rhomb contains too many cooperators so that the condition is not
fulfilled anymore. In view of this result, we find it convenient to 
introduce a parameter to measure the reward in terms of the cost:
\begin{equation}
\label{delta}
\delta\equiv \frac{\rho}{4cN}.
\end{equation}
With this notation, the prediction for the invasion by a single cooperator
is that it will proceed as long as the fraction of cooperators verifies
$N_c(t)/N \leq \delta$.
$N_c(t)$, the number of cooperators at time $t$, can
be easily determined from the recurrence relation for the growing
rhomb: in case the cooperators increase, a new boundary layer is 
added to the rhomb, and we have $N_c(t)=N_c(t-1)+4t$, which can 
be immediately solved (with initial condition $N_c(0)=1$) to give
$N_c(t)=2t^2+2t+1$. Inserting this result in 
the above condition allows to determine the
maximum growth time for the cluster, {that is $t^* = \max\{t: 2t^2 + 2 t + 1\leq \delta N\}$,
and the fraction of cooperators in the steady state:}
\begin{equation}
\label{fraccion}
\frac{N_c(t^*)}{N}.
\end{equation}
So far, we have seen that when the reward is large enough ($\rho > 4cN$), full cooperation 
sets in, whereas for smaller reward, a cooperator cluster grows
up to a final size that depends on $\delta$. 
Interestingly, when $b/2>c$, the reward mechanism is only needed to 
establish an initial population of cooperators, i.e., the rhomb is
resilient.  To show this, notice that
boundary cooperators observe the defectors that earn
the largest payoff (those with two links to two cooperators) and compare
it with the payoff 
obtained by bulk cooperators; boundary cooperators are linked to 
both and unconditional imitation will lead them to adopt the strategy
of the neighbor with the largest payoff. The condition for the cooperators
to resist re-invasion is then 
\begin{equation}
\label{condicion}
4(b-c) > 2b \iff c < \frac{b}{2}.
\end{equation} Indeed, if after a 
number of time steps we turn off the reward, the rhomb structure 
arising from the evolutionary process cannot be re-invaded by 
defectors, as can be seen from Eq.\ (\ref{condicion}).
In the opposite case, $c>b/2$, the reward must be kept at all times
to stabilize
the cooperator cluster.

\begin{figure}
\begin{center}
\includegraphics[width=9cm,height=9cm]{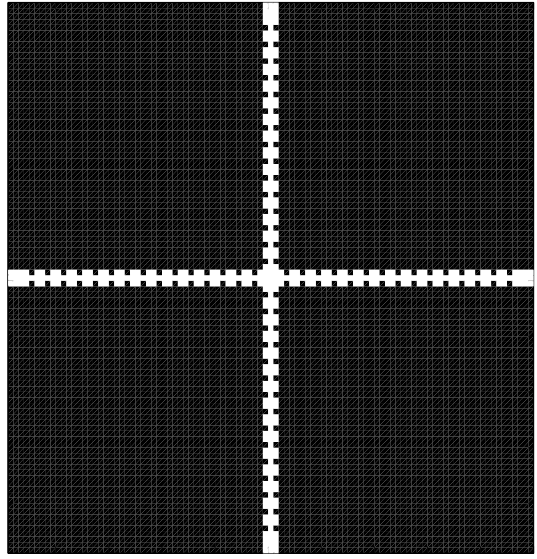}
\end{center}
\caption{Final stage of the invasion of a cooperator population 
by a single defector, for the medium cost case $b/4 < c < b/2$.
Defectors are white, cooperators are black.
\label{figx}}
\end{figure}

{In order to study the resilience of clusters of cooperators,
we consider the simplest case of invasion by a single defector in the Prisoner's Dilemma (without
reward). It can be easily shown
that this  leads to three different cost regimes \citep{maxy:preprint}}:
\begin{itemize}
\item Low cost case, $c < b/4$: the defector is only able to invade its 4 
neighbors, giving rise to a 5 defector rhomb.
\item Medium cost case, $b/4 < c < b/2$: a structure with the shape of
a cross with sawtooth boundaries is formed, implying a finite density of
defectors in the final state (cf.\ Fig.\ \ref{figx}). 
\item High cost case, $c>b/2$: the system is fully invaded by the 
defector, and cooperators go extinct. 
\end{itemize}

\section{Simulations with an initial concentration of defectors}

\begin{figure}
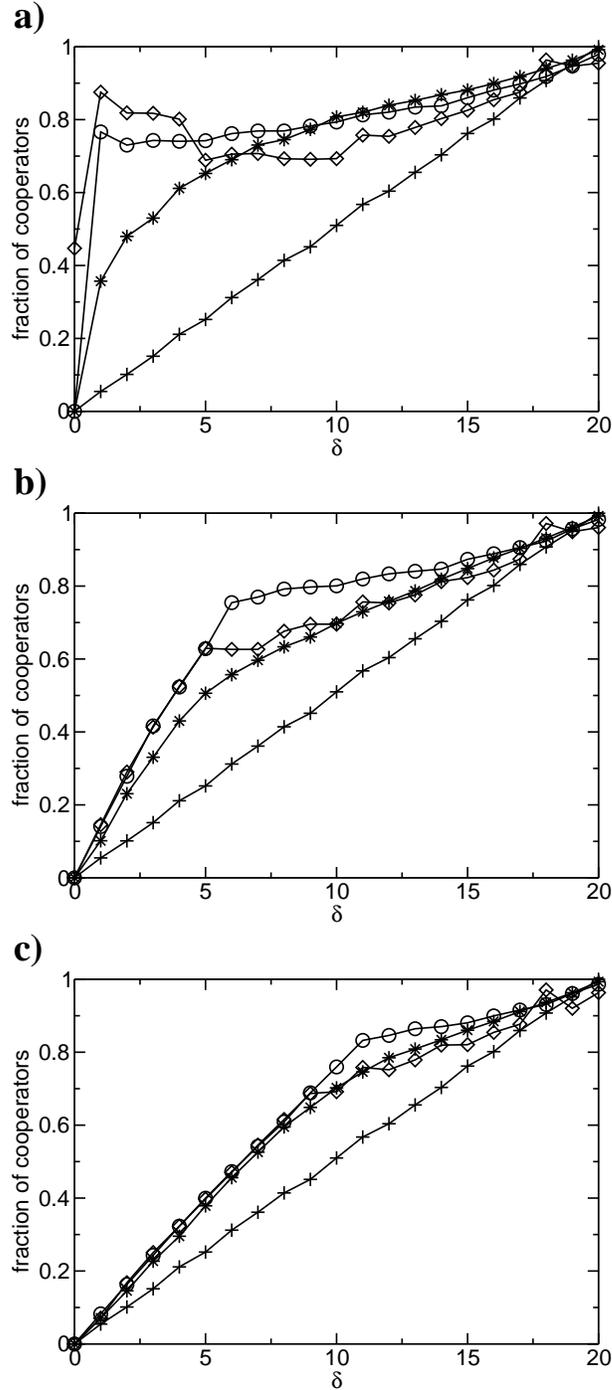

\hspace*{3cm}{\sffamily \large\bf a)}
\begin{center}
\includegraphics[width=8cm,clip=]{c02.eps}\\
\end{center}
{\hspace*{3cm}\sffamily \large\bf b)}
\begin{center}
\includegraphics[width=8cm,clip=]{c04.eps}\\
\end{center}
{\hspace*{3cm}\sffamily \large\bf c)}
\begin{center}
\includegraphics[width=8cm,clip=]{c07.eps}
\end{center}
\caption{Average fraction of cooperators in the steady state as a function of
the rescaled reward $\delta=\rho/4Nc$, obtained starting with 
1 ($+$), 10  ($*$), 100  ($\circ$), and 1000 ($\diamond$) initial cooperators.
a) low cost, $c = 0.2$; b) medium cost, $c = 0.4$;
c), high cost, $c = 0.7$.
\label{fig1}}
\end{figure}

After considering the case of the invasion of a defecting population 
by a single cooperator, we now proceed to a more general situation in 
which there appear a number of cooperators randomly distributed on
the lattice. To this end, we have carried out simulations on square
lattices of size $N=100\times 100$ for 
different initial numbers of cooperators as a function of the 
cost parameter (we take $b=1$ for reference) and the reward. 
A single simulation 
consists of running the game until a steady state is reached, as 
shown by the fraction of cooperators becoming approximately constant.
Generally speaking, the steady state is reached in some 100 games 
per player. For every choice of
parameters, we compute an average over 100 realizations of the initial
distribution of the cooperators. Results are shown in Fig.\ \ref{fig1}
for low, medium and high costs. 

Figure \ref{fig1} shows a number of remarkable features. To begin with, 
the case of invation by a single cooperator reproduces the analytical 
result (\ref{fraccion}), 
On the other hand, in 
all three plots we see that if instead of a single cooperator we have
an initial density of cooperators, the resulting level of cooperation 
is quite higher, particularly when costs are low. Indeed, by looking
at panel a), for which $c=0.2$ ($b=1$), we see that with a 10\%
of initial cooperators cooperation sets in even without reward, as 
observed by \citet{nowak-may}. Notwithstanding, a more remarkable 
result is the fact that with an initial density of cooperators
as low as 0.1\% we find large cooperation levels for small rewards, for 
all values of costs. Clearly, the cooperation level decreases with
increasing cost, but even for high costs [panel c), $c=0.7$], 
the cooperation level is significantly higher than the single
cooperator one. In this last case, we also observe that the 
final state becomes practically 
independent of the density of initial cooperators. Finally, an 
intriguing result is that in the low cost case, the observed
cooperation fraction is not a monotonically increasing function 
of the reward: As it can be seen from the plot, for moderate and
particularly for large initial densities of cooperators, 
increasing the reward may lead to lower levels of cooperation. 
The reason for this phenomenon is that, if the reward increases, 
the cooperator clusters arising from the cooperator invaders grow
larger and overlap.  Therefore, clusters with  
rugged boundaries are formed, allowing for defectors with three 
cooperators which may then be able to reinvade. Further increments
of the reward restore the cooperation levels because then even
these special defectors are overrode. The important 
consequence is that one cannot assume that, for any situation, 
increasing the reward leads to an increasing of the cooperation,
i.e., one has to be careful in designing the reward for each 
specific application. 

\begin{figure}
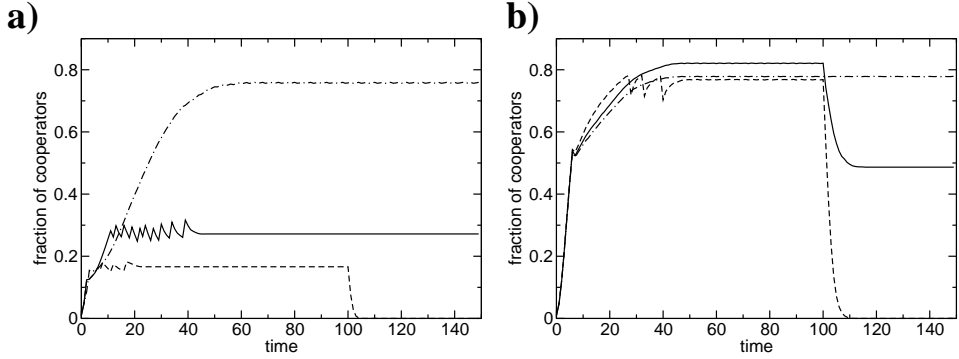

\begin{center}
\hspace*{-6cm}
{\sffamily \large\bf a)} \hspace*{6cm}
{\sffamily \large\bf b)}\\
\includegraphics[width=6cm,clip=]{robust1-color.eps}\ \ \
\includegraphics[width=6cm,clip=]{robust5-color.eps}
\end{center}
\caption{Time evolution of the fraction of cooperators for
the cases of low (dot-dashed line, $c=0.2$), medium (solid line, $c=0.4$) 
and high (dashed line, $c=0.7$) costs, for simulations starting with
100 initial cooperators (density, 1\%) randomly distributed. 
Shown are the cases of a) low ($\delta=0.1$) and b) high
($\delta=0.5$) rewards. Reward is set in place until
$t=100$ and turned off afterwards.
\label{fig2}}
\end{figure}

The other relevant issue to address in the simulations is the 
resilience of the attained cooperation levels. Figure 2 summarizes
our results in this regard. Both for low and high rewards, we 
confirm the result for the single cooperator invasion that 
cooperation disappears if the reward is turned off when 
the costs are high ($c=0.7>b/2$). For moderate and low costs,
the structures arising from the evolution with reward do show
resilience, at least to some degree. Interestingly, the case
of low reward [panel a)] gives rise to extremely robust 
cooperation levels, whereas higher rewards [panel b)] lead
to structures for which cooperation decreases when the reward
is absent (medium cost case). This result is connected with 
the one already discussed that the cooperation level may not 
be monotonics in the reward, and makes it clear that structures
originating from a very agressive, high reward policy may be 
less resilient than those built with low rewards. 

\begin{figure}
{\sffamily \large\bf a)} \hspace*{5.6cm}
{\sffamily \large\bf b)}\\
\includegraphics[width=6cm,height=6cm]{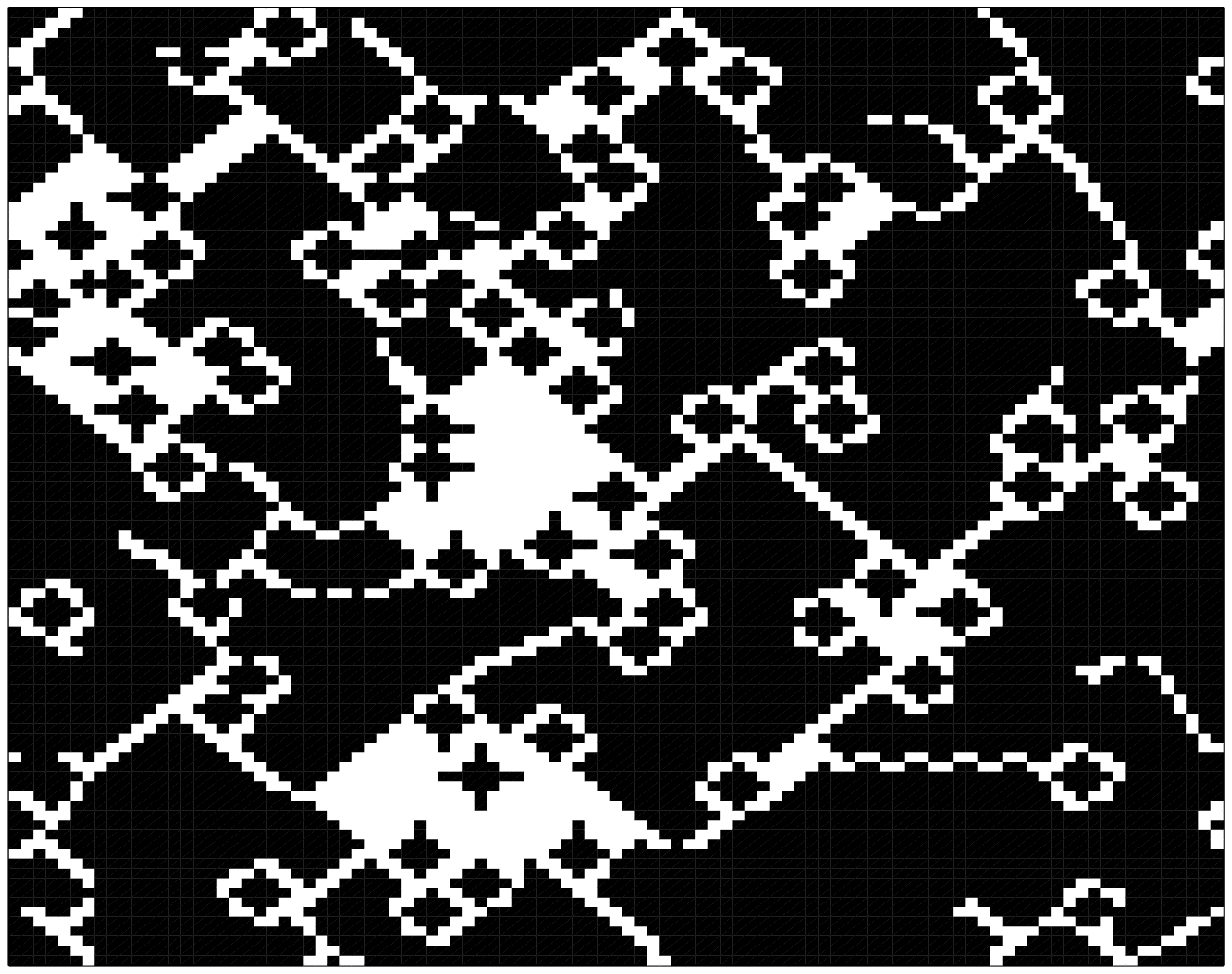}
\includegraphics[width=6cm,height=6cm]{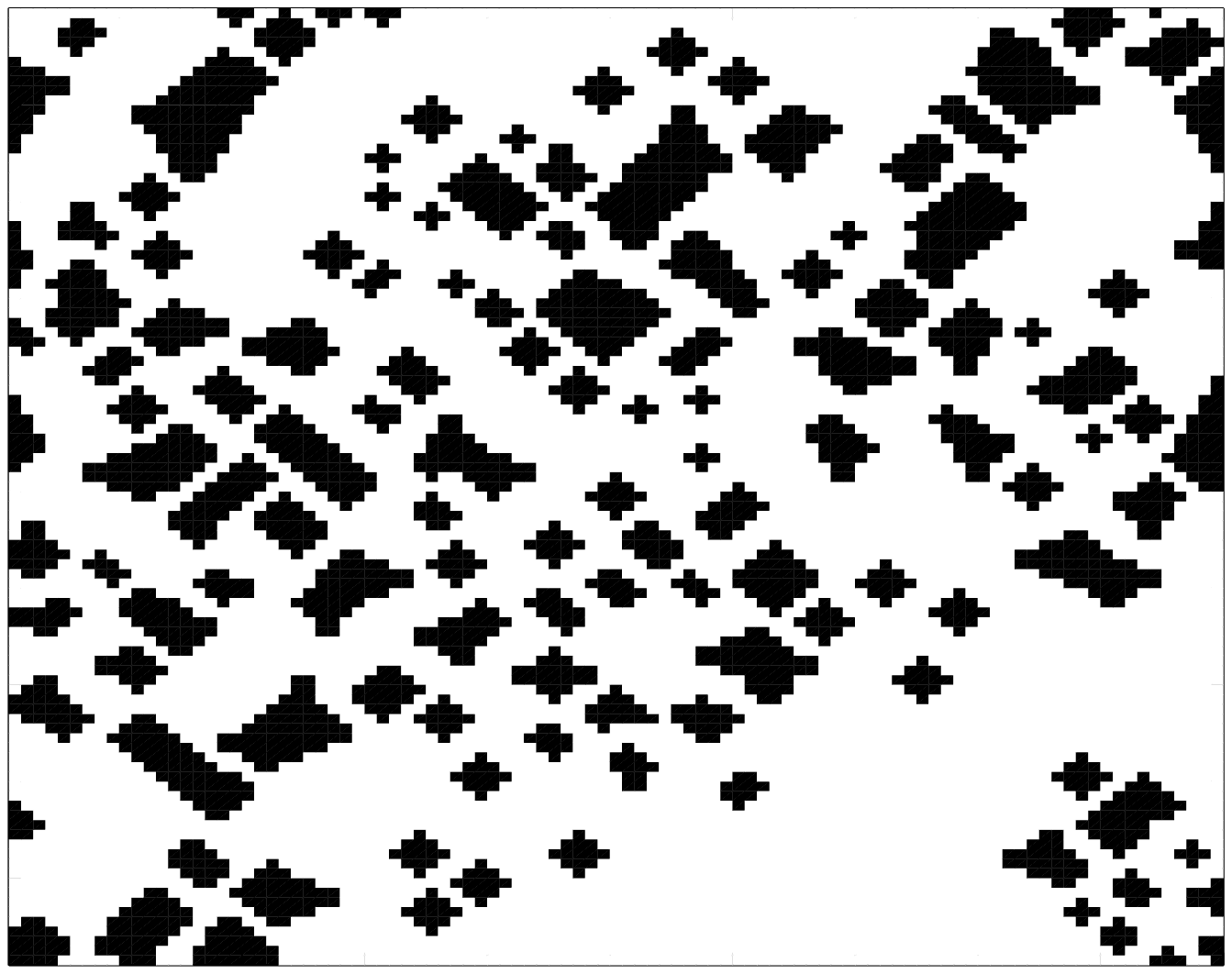}
\caption{System snapshots at the stationary state of a single realization
of the evolution 
(before switching off the reward, see Fig.\ \ref{fig2}) for 
the low reward case ($\delta=0.1$). The initial density of 
cooperators is 1\%. a) low cost ($c=0.2$), b) medium cost 
($c=0.4$). Defectors are white, cooperators are black.
\label{fig3}}
\end{figure}
\begin{figure}
{\sffamily \large\bf a)} \hspace*{5.6cm}
{\sffamily \large\bf b)}\\
\includegraphics[width=6cm,height=6cm]{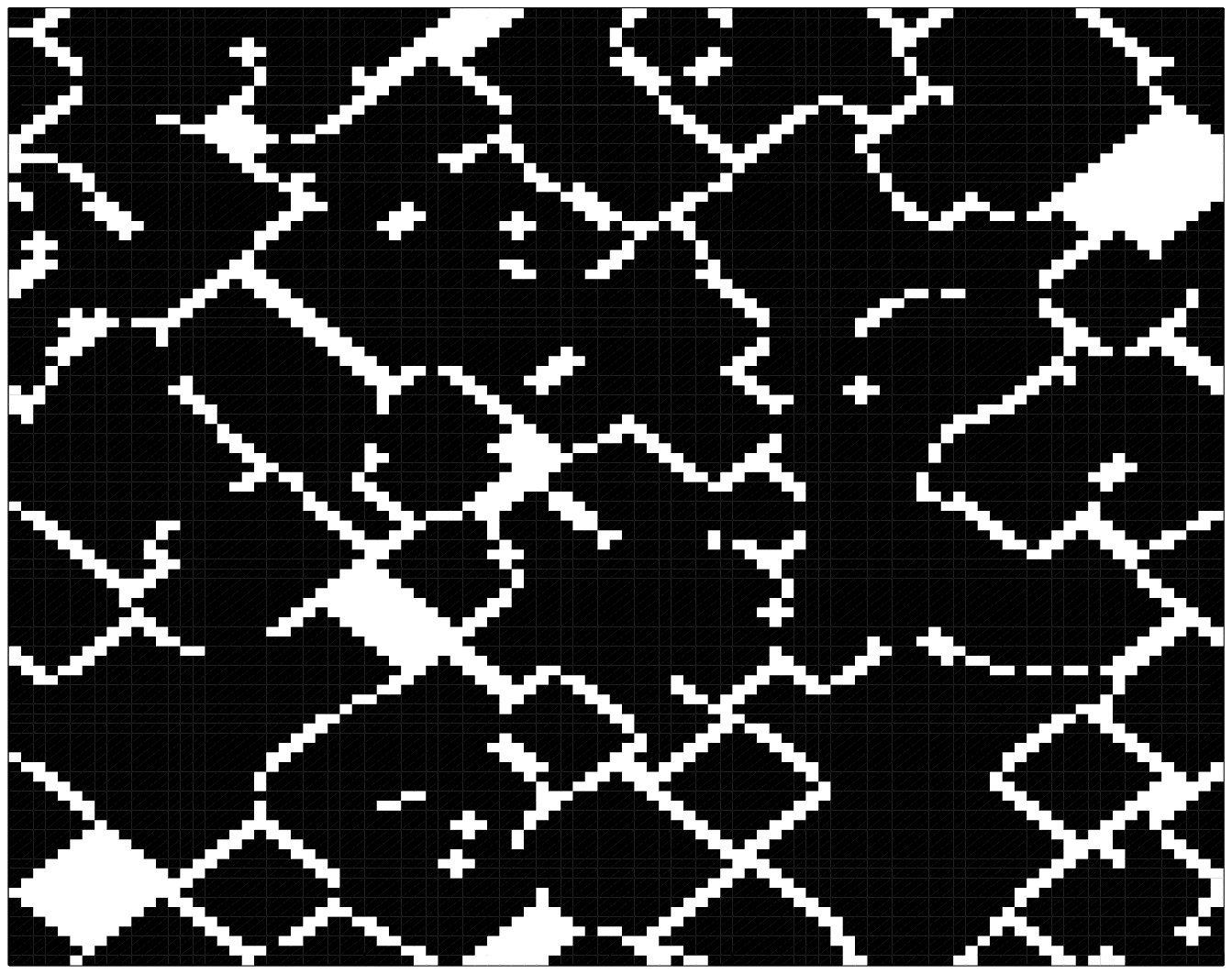}
\includegraphics[width=6cm,height=6cm]{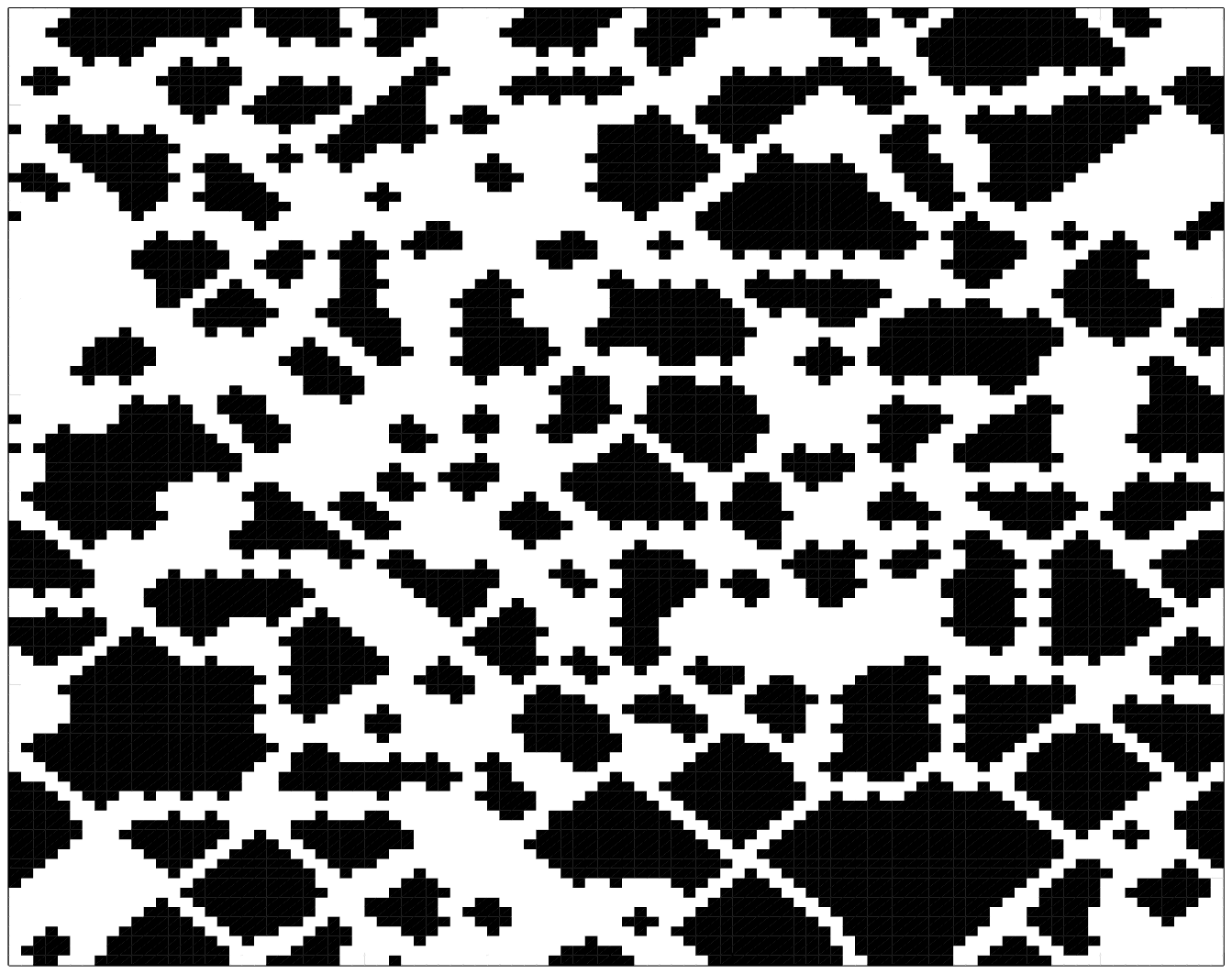}\caption{System snapshots at the stationary state of a single realization
of the evolution, a) before and b) after 
switching off the reward, see Fig.\ \ref{fig2}) for 
the high reward case ($\delta=0.5$) and medium cost 
($c=0.4$). Defectors are white, cooperators are black.
\label{fig4}}
\end{figure}

Further insight on the cluster structure arising from the 
invasion process fueled by the reward can be gained from 
Figs.\ \ref{fig3} and \ref{fig4}. Figure \ref{fig3} shows the
stationary structure of the cooperator clusters for the low
reward case ($\delta=0.1$). As we are now considering that the
initial configuration contains a 1\% of cooperators randomly
distributed, the shapes are irregular, and some rhombs are larger
than others because they merge during evolution. In accordance 
with Fig.\ \ref{fig2}, in the low cost situation the cooperation
level reached is much larger than in the medium cost case. However,
both structures are resilient and survive unchanged if the reward 
is removed. This is due to the fact that, as discussed above, in 
that case defectors can never invade a cooperating population. 
The final structure for the high cost case is similar to Fig.\ 
\ref{fig3}b), but in this case suppression of the reward leads
to an immediate invasion by defectors until they occupy the whole
system. 
When the reward is larger, the situation is somewhat different, as
can be appreciated from Fig.\ \ref{fig4}. While for low cost we again
obtain resilient structures that are preserved even without reward, in
the medium cost regime the patterns change. Panel a) shows the stationary
state reached with the reward; when the reward is taken away, the state
changes and evolves to the configuration shown in panel b). What is 
taking place here is that due to the high reward, a cooperation level
close to 1 is reached, most of the defectors being isolated or along
lines. When the reward is switched off, these defectors are in a
position to rip much payoff from their interactions with the cooperators,
allowing for a partial reinvasion. Therefore, the final cooperation level
has more or less halved. We stress that even then the cooperation level
that remains after the suppression of the reward is rather large 
(about 60\%), another hint of the efficiency of this mechanism to 
promote cooperation.  

\section{Discussion and conclusions}

We have proposed a mechanism that allows a population of 
cooperators to grow and reach sizeable proportions in the 
spatial Prisoner's Dilemma in a cost-benefit framework. 
This mechanism is based in the
distribution of a fixed-amount reward among all cooperators
at every time step. With this contribution to the payoffs
of the standard Prisoner's Dilemma, even a single cooperator
is able to invade a fully defecting population. The 
resulting cooperator fraction is determined by the amount
of the reward as compared to the total number of players 
and to the cost of the interaction. Furthermore, for low
and medium costs ($c<b/2$) cooperation is resilient in 
the sense that if at some time step the reward is suppressed,
the cooperator cluster cannot be re-invaded by the defectors. 
Finally, we have seen that low rewards are capable to induce
a very large cooperation level, so the mechanism works even 
when it changes only a little the payoffs of the Prisoner's 
dilemma. 

The result we have obtained is relevant, in the first place,
as a necessary complement of the original work by \citet{nowak-may}
{within the cost-benefit context}. 
In their work they showed that the spatial structure allowed
cooperator clusters to survive and resist invasion by defectors,
but they began with a large population of cooperators. Our work
provides a putative explanation as to where this population  
comes from. We note that in the original work by \citet{nowak-may}
they observed that the cooperation level decreased with respect
to the initial population, so a mechanism leading to the 
appearance of high cooperator levels is certainly needed. 
In this regard, we want to stress that the reward mechanism 
gives rise to structures with very good resilience properties:
Simulations without the reward show that starting from 
a randomly distributed population of cooperators 
with very large density ($\sim 90\%$), the final cooperation 
level is halved for low costs, and practically disappears
for moderate costs. 

%
We stress that, to our knowledge, this is the first time that a
mechanism based on a fixed-amount reward to be shared among 
cooperators is proposed. Notwithstanding, there are other proposals
which are somewhat related to ours, most prominent among them 
being those by \citet{lugo} and \citet{hauert2}. 
\citet{lugo} introduce a tax mechanism in which everybody in the
population contributes towards a pool that is subsequently distributed
among the cooperators. This is different from the present proposal
in so far as the contribution from the tax is not a fixed 
quantity but rather it increases with the average payoff. 
On the other hand, \citet{hauert2} focuses 
on the effects of nonlinear discounts (or synergistic 
enhancement) depending on the number of cooperators in the
groups of interacting individuals. Although the
corresponding game theoretical model, discussed by \citet{hauert1} 
belongs to the same general class of $n$-player games of our shared
reward model, the spatial implementation of the two models is very
different. Thus, in \citet{hauert2}, payoffs for a given individual
depend on the number of cooperators in her neighborhood, whereas in 
the present work payoffs depend on the total number of cooperators in
the network. On the other hand, our
interest is also different, in so far as we are discussing 
a mechanism to foster the appearance of an initial, sizeable
population of cooperators which can later be stable without
this additional resource. It is important to stress that with 
our mechanism a large level of cooperation can be established
and (in the appropriate parameter range) stabilized. 

We believe that our results may be relevant for a number
of experimental situations where the Prisoner's Dilemma has been
shown to appear in nature. Thus, the stabilization of mutualistic
symbioses by rewards or sanctions as observed in, e.g., legume-rhizobium 
mutualism \citep{denison} is related to the mechanism we
are proposing here: It is observed that soybeans penalize 
rhizobia that fail to fix $N_2$ in their root nodules. This decreases
the defector's payoff which is similar to increasing the 
cooperator's payoff by a reward. On the other hand, a description
of the interaction between different strains of microorganisms 
[see \citet{crespi,velicer} and references therein] in terms of
this reward mechanism instead of the standard Prisoner's Dilemma
may prove more accurate and closer to the actual interaction 
process. An example could be the evolution of cooperators with 
reduced sensitivity to defectors in the RNA Phage $\Phi$6 
\citep{turner,turner2}.
Cooperative foraging is another context where the mechanism of
rewarding cooperation may be relevant, ranging from microorganisms
such as {\em Myxococcus xanthus} \citep{dworkin} 
through beetles \citep{berryman} to wolves or lions
\cite{anderson}.
Finally, the question arises as to the validity of 
such a mechanism to promote cooperation within humans, as individual
players can not predict in advance the additional payoff they will
obtain from the reward, and therefore it is not clear whether it
would have an influence on them or not. Evidences from cooperative 
hunting in humans \citep{alvard,alvard2} show that high levels of
sharing help sustain cooperative behavior. However, in the human 
case, contexts where the reward would be more explicitly included
in a manner transparent to the players are possible and amenable 
to experiments. Research along these 
lines is necessary to assess the possible role of the reward 
mechanism in specific situations.

\section*{Acknowledgments}

This
work is partially supported by Ministerio de Educaci\'on y Ciencia (Spain) under grants
Ingenio-MATHEMATICA, MOSAICO and NAN2004-9087-C03-03 and by Comunidad de Madrid
(Spain) under grants
SIMUMAT and MOSSNOHO.





\begin{thebibliography}{}


\harvarditem{Alvard}{2001}{alvard} Alvard, M. 2001. Mutualistic hunting. 
In {\em The Early Human Diet: The Role of
Meat}, Craig Stanford and Henry Bunn, eds., pp. 261-278.  Oxford
University, New York. 

\harvarditem{Alvard2}{2003}{alvard2} Alvard, M. 2004, Good hunters keep smaller 
shares of larger pies. Behav.\ Brain Sci.\ 27, 560--561. 

\harvarditem{Anderson and Franks}{2001}{anderson} Anderson, C., Franks, N.\ R., 2001. Teams in animal societies. Behav.\ Ecol.\ 12, 534--540.

\harvarditem{Axelrod}{1984}{axelbook} Axelrod, R., 1984. The Evolution of Cooperation. Penguin, London.

\harvarditem{Axelrod and Hamilton}{1981}{axelrod} 	Axelrod, R., Hamilton, W.\ D., 
1981. The evolution of cooperation. Science 211, 1390--1396. 

\harvarditem{Berryman {\em et al.}}{1985}{berryman}  Berryman, A.\ A., 
Dennis, B., Raffa, K.\ F., Stenseth, N. C., 1985. Evolution of optima 
group attack with particular reference to bark beetles (Coleoptera:Scolytidae).
Ecology 66, 898--903.

\harvarditem{Camerer}{2003}{camerer} Camerer, C., 2003. Behavioral Game
Theory: Experiments in Strategic Interaction. Princeton University Press,
NJ.

\harvarditem{Connor}{1995}{connor} Connor, R. C., 1995. The benefits of 
mutualism: a conceptual framework. Biol.\ Rev.\ 70, 427--457.

\harvarditem{Crespi}{2001}{crespi} Crespi, B. J., 2001. The evolution of
social behavior in microorganisms. Trends Ecol.\ Evol.\ 16, 178--183.

\harvarditem{Cuesta {\em et al.}}{2007}{nosotros} Cuesta, J.\ A., Jim\'enez, R., Lugo, H., 
 S\'anchez A., 2007. Rewarding cooperation in social dilemmas. Working paper 
 07--52, Departamento de Econom\'\i a, Universidad Carlos III de Madrid.

\harvarditem{Doebeli and Hauert}{2005}{doebeli} Doebeli, M., Hauert, C., 2005. Models of cooperation based on
the Prisoner's Dilemma and the Snowdrift game. Ecol.\ Lett.\ 8, 748--766.

\harvarditem{Dugatkin {\em et al.}}{1992}{dugatkinold} Dugatkin, L.\ 
A., Mesterton-Gibbons,  Houston, A.\ I., 1992. Beyond the prisoner's dilemma:
Toward models to discrimate among mechanisms of cooperation in nature. 
Trends Ecol.\ Evol.\ 7, 202--205. 

\harvarditem{Dugatkin and Mesterton-Gibbons}{1996}{dugatkin} Dugatkin, L.\ 
A., Mesterton-Gibbons, M., 1996. Cooperation among unrelated individuals: 
reciprocal altruism, by-product mutualism and group selection in fishes. 
BioSystems 37, 19--30.

\harvarditem{Dworkin}{1996}{dworkin}  Dworkin, M., 1996. Recent advances in
the social and developmental biology of the Myxobacteria. Microbiol.\ 
Rev.\ 60, 70--102. 

\harvarditem{Fehr and G\"achter}{2002}{fehr}  Fehr, E., G\"achter, S., 2002.
Altruistic punishment in humans. Nature 415, 137--140. 

\harvarditem{Frick and Schuster}{2003}{frick} Frick, T., Schuster, S., 2003.
An example of the prisoner's dilemma in biochemistry. Naturwiss.\ 90, 327--331.

\harvarditem{Gintis}{2000}{gintis} Gintis, H., 2000. Game Theory Evolving. Princeton
University Press, Princeton, NJ.

\harvarditem{Hamilton}{1964}{hamilton} Hamilton, W. D., 1964. The genetical 
evolution of social behavior I. J.\ Theor.\ Biol.\ 7, 1--16. 

\harvarditem{Hammerstein}{2003}{hammerstein} Hammerstein, P.\ (Ed.), 2003. Genetic and Cultural Evolution of 
Cooperation. MIT Press, Cambridge, MA. 

\harvarditem{Hardin}{1968}{hardin}
Hardin, G., 1968. The tragedy of the commons. Science 162, 1243--1248.

\harvarditem{Hauert}{2006}{hauert2} Hauert, C., 2006. Spatial effects in 
social dilemmas. J.\ Theor.\ Biol.\ 240, 627--636.

\harvarditem{Hauert {\em et al.}}{2006}{hauert1} Hauert, C., Michor, F.,
Nowak, M. A., Doebeli, M., 2006. Synergy and discounting of cooperation in
social dilemmas. J.\ Theor.\ Biol.\ 239, 195--202.  

\harvarditem{Jim\'enez {\em et al.}}{2007}{maxy:preprint} Jim\'enez, R., Lugo, H., 
Egu\'{i}luz, V., San Miguel, M., 2007. Learning to cooperate. Unpublished manuscript. 

\harvarditem{Kiers {\em et al.}}{2003}{denison} Kiers, E.\ T., Rousseau, R.\ A., West,
S.\ A., Denison, R.\ F., 2003. Host sanctions and the legume-rhizobium 
mutualism. Nature 425, 78--81.

\harvarditem{Kollock}{1998}{dilemas} Kollock, P., 1998. Social dilemmas: The
anatomy of cooperation. Annu.\ Rev.\ Sociol.\ 24, 183--214.

\harvarditem{Legge}{1996}{legge} Legge, S, 1996. Cooperative lions escape the Prisoner's Dilemma. Trends Ecol.\ Evol.\ 11, 2--3.

\harvarditem{Lugo and Jim\'enez}{2006}{lugo} Lugo, H., Jim\'enez, R., 2006.
Incentives to cooperate in network formation. Comp.\ Econ.\ 28, 15--27.

\harvarditem{Maynard-Smith}{1982}{maynard} Maynard-Smith, J., 1982. Evolution
and the Theory of Games. Cambridge University Press, UK.

\harvarditem{Nowak}{2006}{nowak-book} Nowak, M.\ A., 2006. Five rules for 
the evolution of cooperation. Science 314, 1560--1563.

\harvarditem{Nowak}{2006}{nowak-review} Nowak, M.\ A., 2006. Evolutionary 
Dynamics. Harvard University Press, Cambridge, MA.

\harvarditem{Nowak and May}{1992}{nowak-may} Nowak, M.\ A.\, May, R.\ M.\ May, 1992. Evolutionary
games and spatial chaos. Nature 415, 424--426.

\harvarditem{Nowak and Sigmund}{1998}{nowak-sigmund2} Nowak, M.\ A., Sigmund, K.,
1998. Evolution of indirect reciprocity by image scoring. Nature 393, 
573--577. 

\harvarditem{Nowak and Sigmund}{2004}{nowak-sigmund} Nowak, M.\ A., Sigmund, K., 2004. Evolutionary
dynamics of biological games. Science 303, 793--799.

\harvarditem{Ohtsuki {\em et al.}}{2006}{japo} Ohtsuki, H., Hauert, C., 
Lieberman, E., Nowak, M.\ A., 2006. A simple rule for the evolution of cooperation on graphs and social networks. Nature 441, 502--505.

\harvarditem{Packer and Ruttan}{1988}{packer} Packer, C., Ruttan, L., 1988. The evolution of cooperative hunting. Am.\ Nat.\ 132, 159--198.

\harvarditem{Pennisi}{2005}{pennisi} Pennisi, E., 2005. How did cooperative behavior evolve? 
Science 309, 93. 

\harvarditem{Samuelson}{1954}{samuelson} Samuelson, P.\ A., 1954.
The pure theory of public expenditure. 
Rev.\ Econ.\ Stat.\ 36, 387--389.

\harvarditem{Skyrms}{2004}{skyrms} Skyrms, B, 2006. The Stag Hunt and 
the Evolution of
Social Structure. Cambridge University Press, Cambridge, UK. 

\harvarditem{Szab\'o and F\'ath}{2007}{szabo} Szab\'o, G., F\'ath, G., 2007. 
Evolutionary games on graphs. Phys.\ Rep., in press. 

\harvarditem{Trivers}{1971}{trivers} Trivers, R.\ L., 1971. The evolution 
of reciprocal altruism. Q.\ Rev.\ Biol.\ 46, 35--57.

\harvarditem{Turner and Chao}{1999}{turner} Turner, P.\ E., Chao, L., 1999. 
Prisoner's dilemma in an RNA virus. Nature 398, 441--443. 

\harvarditem{Turner and Chao}{2003}{turner2} Turner, P.\ E., Chao, L., 2003. 
Escape from prisoner's dilemma in RNA phage $\Phi$6. Am.\ Sci.\ 161, 497--505.

\harvarditem{Velicer}{2003}{velicer} Velicer, G.\ J., 2003. Social strife in the microbial world. 
Trends in Microbiol.\ 11, 330--337.

\harvarditem{Vuli\'{c} and Kolter}{2001}{vulic} Vuli\' {c}, M., Kolter, R., 
2001. Evolutionary cheating in {\em Escherichia coli} stationary phase 
cultures. Genetics 158, 519--526. 

\harvarditem{Wingreen and Levin}{2006}{levin} Wingreen, N.\ S., Levin, S.\ A.,
2006. Cooperation among microorganisms. PLOS Biology 4, 1486--1488.
\end{thebibliography}
\end{document}